\documentclass[11pt,twoside]{article}

\usepackage{asp2004}
\usepackage{epsf}
\usepackage{psfig}
\usepackage{lscape}
\usepackage{graphicx}

\begin{document}
\title{Young, Jupiter-Mass Objects in Ophiuchus}   
\author {K. N. Allers, D. T. Jaffe}
\affil{Department of Astronomy,
University of Texas at Austin, Austin, TX 78712-0259}
\author {N. S. van der Bliek}
\affil{Cerro Tololo Inter-American Observatory, Casilla 603, La Serena, Chile}
\author{F. Allard, I. Baraffe}  
\affil{Centre de Recherche Astronomique de Lyon (UML 5574), \'{E}cole Normale 
Sup\'{e}rieure, 69364 Lyon Cedex 07, France}

\begin{abstract} 
We have used 3.5 to 8 $\mu$m data from the Cores to Disks (c2d) Legacy survey 
and our own deep IJHKs images of a 0.5 square degree portion of the c2d 
fields in Ophiuchus to produce a sample of candidate young objects with 
probable masses between 1 and 10 M$_{\rm Jupiter}$.  
The availability of photometry over 
whole range where these objects emit allows us to discriminate between young, 
extremely low-mass candidates and more massive foreground and background 
objects and means our survey will have fewer false 
positives than existing near-IR surveys.  The sensitive inventory of a star 
forming cloud from the red to the mid-IR will allow us to constrain the IMF 
for these non-clustered star formation regions to well below the deuterium 
burning limit.  For stars with fluxes in the broad gap between the 2MASS 
limits and our limits, our data will provide information about the 
photospheres.  We will use the Spitzer results in combination with current 
disk models to learn about the presence and nature of circumstellar disks 
around young brown dwarfs.
\end{abstract}
\section{Introduction}   
Free-floating objects with masses comparable to the masses of the most 
substantial extrasolar planets have been difficult to find and even more 
difficult to confirm.  Several groups have reported sources with masses 
below 10 M$_{\rm J}$ \citep{martin04,lucas03}.  
The limited wavelength range of the 
photometry available for these objects makes the inital source identifications 
uncertain.  All of the candidate objects lie at distances $\sim$450 pc, where 
their extreme faintness makes them difficult to confirm spectroscopically.  
In fact, the only spectrospically ``confirmed'' extremely low mass source 
from one of these samples \citep{zapatero02} may be an older foreground 
object with higher mass \citep{burgasser04}.  It is well worth continuing to 
search to search for a sample of young objects with extremely low masses, 
both because of the clues they provide about star and planet formation and 
because they could serve as a testbed for ideas about the structure and early 
evolution of massive planets.

\section{Experimental Design}
The problem of finding Jupiter-mass objects in nearby molecular clouds is 
more one of 
identification than detection.  Models for very young stellar objects predict 
that even extremely low mass bodies are within the reach of direct 
observations in the near and mid-IR \citep{baraffe03,burrows01}.  
\begin{figure}[!ht]
\begin{center}
\includegraphics[scale=0.4, clip=true]{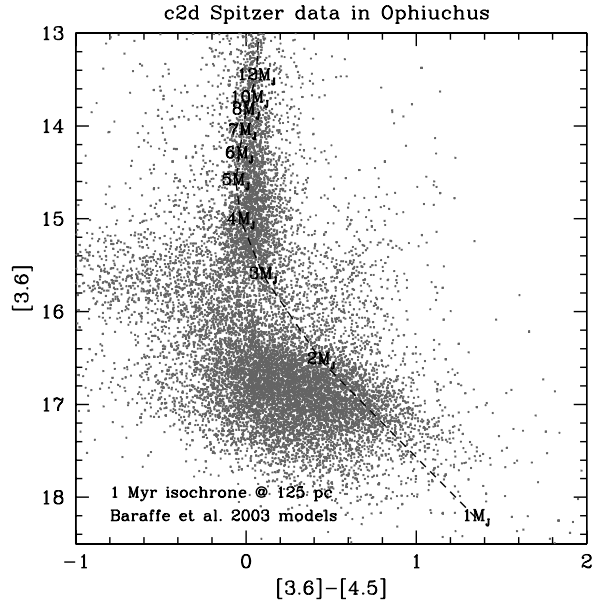}
\caption{Color magnitude diagram for IRAC bands 1 and 2 observed as a part of 
the c2d survey towards $\sim$0.5 square degrees in Ophiuchus.  
The diagram contains $\sim$18,000 sources.  
The dashed line is the 10$^6$ year isochrone from the \citet{baraffe03} 
models.  The data points in the upper part of the diagram are primarily 
background stars, while those in the lower part are largely galaxies.}
\end{center}
\end{figure}
The Spitzer Legacy Program ``From Molecular Cores to Planet-Forming Disks'' 
\citep[c2d][]{evans03} provides mid-IR fluxes (in the [3.6 and [4.5] micron 
bands) for objects with theoretical masses down to 2 M$_{\rm J}$.  Figure 1 
illustrates, however, that using IRAC colors alone, one cannot distinguish 
young, low-mass objects from background stars and galaxies.
Our survey of part of the Ophiuchus cloud in I,J,H, and Ks using MOSAIC II 
and ISPI on the Blanco 4m telescope has 10$\sigma$ limits of I=23.5, J=20, 
H=19, and Ks=18.5.  These limits allow us (based on theoretical isochrones) to 
detect 2 M$_{\rm J}$, 10$^6$ year old objects even in the presence of modest 
extinction (A$_{\rm V}<$10).  The fluxes between 0.8 and 3.5 $\mu$m, where our 
survey should be complete for all 2 M$_{\rm J}$ sources and where most of the 
flux 
from these sources emerges, provide us with a way to build enough 
color-color and color-magnitude spaces to break the degeneracy between 
our target population and the myriad of contaminants.

\section{A Sample of Candidate Young, Jupiter-Mass Objects}
In our first-round analysis of $\sim$0.5 sq degrees in Ophiuchus, 
we start with 
19,000 objects detected at $>5\sigma$ in all 5 bands used for our cuts: 
I,J,H,Ks, and [3.6].  In our current pass through the sample, we use a 
set of empirical criteria based on the 
nominal colors and magnitudes of a 10 M$_{\rm J}$, 10$^6$ year old object
\citep{chabrier00}.  We 
eliminate all sources with J$<15.09$ and I-J$<2.94$, 
thereby removing sources that are too 
bright either because they are foreground objects, luminous background 
objects, or galaxies with blue I-J 
colors.  Of the remaining 6,000 sources that are faint in J and 
red in I-J, most 
are reddened background M stars.  We look at this reduced sample in the 
IJH and IJK 
color planes and deredden all sources back to the theoretical main sequence 
for 10$^6$ year old objects.  At this point, only 50 objects have I-J$>2.94$.  
Since brown dwarfs get monotonically redder in K-L as spectral types get later 
\citep{golimowski04}, we cut the sample further by requiring that dereddened 
K-[3.6]$>0.46$, leaving us with 37 total candidate 10$^6$ year old, 1 to 10 
M$_{\rm J}$ objects.  Figure 2 shows the observed colors of one of our 
candidate objects.  Though the colors agree quite well with model predictions 
for a young 2 M$_{\rm J}$ object, higher-mass late M and early L type 
field brown dwarfs have similar colors \citep{patten04}.

\begin{figure}[!ht]
\begin{center}
\includegraphics[scale=0.4, clip=true]{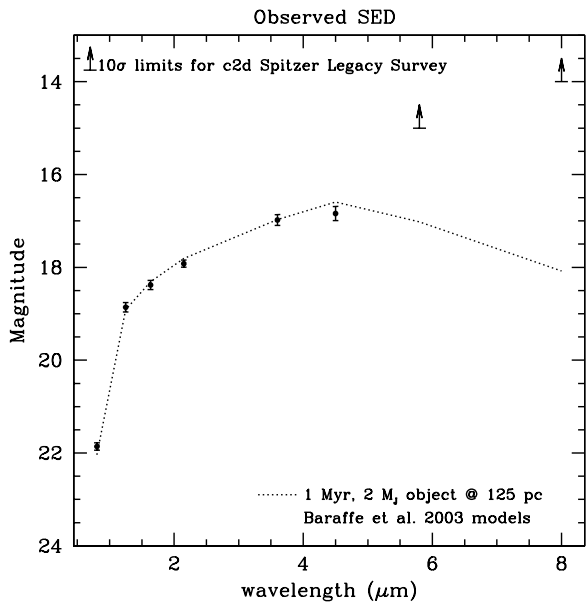}
\caption{Observed 0.8 to 4.5 $\mu$m fluxes (points with error bars) and longer 
wavelength upper limits for a candidate young, low-mass object in Ophiuchus.  
Superposed is a model for a 10$^6$ year old, 2 M$_{\rm Jupiter}$ object 
\citep{baraffe03,allard01}.}
\end{center}
\end{figure}

Ultimately, we need spectroscopy to confirm that our candidate objects have 
low gravities, and therefore have low masses.  The shape of broad H$_2$O 
and CH$_4$ absorption bands in near-IR spectra are sensitive to gravity
\citep{lucas01}, while the relative strengths of the bands can provide 
spectral types \citep{geballe02}.  Once we have spectroscopically confirmed 
objects with masses of 1-10 M$_{\rm J}$, the {\it observed} colors of these 
objects can add confidence to the low-mass nature of other candidate objects 
in our survey.  We will also adjust the models based on the observed spectra 
and colors of our confirmed objects and tighten our selection criteria for 
future passes through our data.

Spectroscopy is not the only way to gain confidence in our selection 
techniques or to produce a subsample with higher reliability.  
Excess emission in the IRAC bands has been 
reported around a spectroscopically confirmed 15 M$_{\rm J}$ object in 
Chamaeleon \citep{luhman05}.  We have 
recently examined our sample of candidates for evidence of excess mid-IR 
emission from circum-object disks.  Most of our candidates are too faint 
to be detected in IRAC bands 3 or 4 of the c2d survey, 
even if they have excess emission from a disk.
Among the 37 candidates in our sample, 5 show 
excess emission in [5.8] and/or [8.0] compared to the IRAC colors of field 
brown dwarfs with comparable near-IR colors \citep{patten04}.  2 of the 5 
sources showing mid-IR excess are also dectected at 
24 $\mu$m in the c2d MIPS data.  Preliminary modeling of the SEDs of our 
candidates detected at 24 $\mu$m indicates that these objects, with 
possible masses as low as 5 M$_{\rm J}$, could have flared circum-object disks.
\acknowledgements 
The authors wish to thank Neal Evans and the Cores to Disks Legacy team for 
providing IRAC data for our regions.  We also thank Giovanni Fazio and Brian 
Patten for providing their IRAC results for field brown dwarfs.


\end{document}